\documentclass[prd,twocolumn, showpacs,nofootinbib,superscriptaddress]{revtex4}
\usepackage{graphicx}
\usepackage{here}
\usepackage{framed}
\usepackage{amstext,amsmath,amssymb}
\usepackage{latexsym}

 \def \v{\vspace}
\newcommand{\Ref}[1]{(\ref{#1})}

  \def\cc{{\cal C}} \def\dd{{\cal D}}

\def\hh{{\cal H}}   
\def\lll{{\cal L}}  \def\nn{{\cal N}} 
   \def\ss{{\cal S}}

\def\be{\begin{equation}} \def\ee{\end{equation}}
\def\bes{\begin{eqnarray}} \def\ees{\end{eqnarray}}
\newcommand{\beq}{\begin{equation}} \newcommand{\eeq}{\end{equation}}
\newcommand{\beqa}{\begin{eqnarray}} \newcommand{\eeqa}{\end{eqnarray}}

\def\nn{\nonumber} \def\arr{\rightarrow}

\def\w{\wedge} \def\ka{\kappa}
 
\def\f{\frac}
\def\tl{\widetilde} 

 \def \vphi{\varphi}
 
\def\hh{{\cal H}}

 \newcommand{\lalg}[1]{\mathfrak{#1}}
 \newcommand{\SO}{\mathrm{SO}}

\newcommand{\so}{\lalg{so}} 
 
\renewcommand{\v}{\overrightarrow}

\newcommand{\pa}{\partial}

\begin{document}

%\title{Mechanics of a Free Particle in Deformed Special Relativity}
\title{The Free Particle in Deformed Special Relativity}

\author{{\bf Florian Girelli}}
\email{fgirelli@perimeterinstitute.ca}
\affiliation{Perimeter Institute, 31 Caroline Street North
Waterloo, Ontario, Canada N2L 2Y5}

\author{{\bf Tomasz Konopka}}
\email{tkonopka@perimeterinstitute.ca} 
\affiliation{Perimeter Institute, 31 Caroline Street North
Waterloo, Ontario, Canada N2L 2Y5}
\affiliation{University of 
Waterloo, 200 University Ave. West, Waterloo, ON, N2L 3G1, 
Canada} 

\author{{\bf Jerzy Kowalski-Glikman}}
\email{jurekk@ift.uni.wroc.pl} 
\affiliation{Institute for 
Theoretical Physics, University of Wroc\l{}aw, Pl.\ Maxa Borna 9, 
Pl--50-204 Wroc\l{}aw, Poland}

\author{{\bf Etera R. Livine}}
\email{elivine@perimeterinstitute.ca}
\affiliation{Perimeter Institute, 31 Caroline Street North
Waterloo, Ontario, Canada N2L 2Y5}
\affiliation{Laboratoire de Physique, ENS Lyon, CNRS UMR 5672, 46 All\'ee d'Italie, 69364 Lyon Cedex 07}

\begin{abstract}
%\begin{center}
%{\small ABSTRACT}
%\end{center}
The phase space of a classical particle in DSR contains de Sitter 
space as the space of momenta. We start from the standard 
relativistic particle in five dimensions with an extra constraint 
and reduce it to four dimensional DSR by imposing appropriate 
gauge fixing. We analyze some physical properties of the 
resulting theories like the equations of motion, the form of 
Lorentz transformations and the issue of velocity. We also 
address the problem of the origin and interpretation of different 
bases in DSR.
\end{abstract}

\maketitle

%%%%%%%%%%%%%%%%%%%%%%%%%%%%%%%%%%%%%%%%%%%%%%%%%%%%%%%%%%%%%%%%%%%%%%%%%%
%%%%%% ARTICLE %%%%%%%%%%%%%%%%%%%%%%%
%%%%%%%%%%%%%%%%%%%%%%%%%%%%%%%%%%%%%%%%%%%%%%%%%%%%%%%%%%%%%%%%%%%%%%%%%%

\section{Introduction}

In many situations, it is of interest to discuss phase spaces 
with curvature. Examples of models with curved position spaces 
abound, but models with curved momentum space can also be quite
interesting 
\cite{snyder,Donkov:1984fj,Donkov:1984ax,Kadyshevsky:1983yc}. One 
of their common features is that they are all closely related to 
non-commutative geometries (see eg.\ \cite{majidbook}). Some new 
investigations suggest that these models might be relevant to 
quantum gravity. For instance, it has been found that a particle 
coupled to gravity in three dimensions has a curved, maximally 
symmetric momentum space \cite{Matschull:1997du,Freidel:2005bb}. 
Similar properties have been argued to hold in four dimensions as 
well \cite{Imilkowska:2005vs}. As another example, Deformed 
Special Relativity (DSR) 
\cite{Amelino-Camelia:2000ge,Amelino-Camelia:2000mn,jkgminl,rbgacjkg,Magueijo:2001cr}, 
which has been argued to represent an effective low-energy 
description of quantum gravity, also implements momentum space as 
de Sitter \cite{Kowalski-Glikman:2002ft,Kowalski-Glikman:2003we} 
or anti de Sitter \cite{Blaut:2003wg}. 

In usual relativistic physics, the phase space of a particle 
consists of the position space and the momentum space both of 
which are four-dimensional Minkowski spaces. One also introduces 
an action which determines the canonical coordinates on the phase 
space and their dynamics. On the other hand, in DSR, kinematics 
and dynamics are usually defined separately. One first specifies 
a deformed symplectic form on the phase space which reflects the 
fact that the momentum space is de Sitter. The dynamics is then 
determined by the symmetries and the Hamiltonian is constructed 
from the Casimir of the Poincar\'e group. The issue that we 
address here is to determine an action principle for DSR which 
generates both the symplectic form and the dynamics.

The next section is devoted to description of the standard 
relativistic particle, with special emphasis put on the way to 
introduce time and gauge fixing.

In section \ref{sec_dsr}, we 
explain how to reduce the relativistic particle in a 
five-dimensional Minkowski space to a particle with a 
four-dimensional curved de Sitter space of momenta. We constrain 
the 10d flat phase space with an appropriate gauge fixing to 
reduce it to a 8d phase space with a de Sitter energy--momentum 
sector. Since our starting point is the phase space with natural 
coordinates in 10 dimensions, choosing a gauge fixing condition 
is equivalent to defining some particular coordinates in the 
resulting 8 dimensional phase space. We take the reduced phase 
space to define the single particle in DSR.

In sections \ref{sec_snyder} and \ref{sec_bcp}, we follow the 
gauge fixing procedure and provide the explicit gauge fixing 
conditions leading to the Snyder and the bicrossproduct bases (we 
deal with these two bases because they are the most frequently 
used DSR models.) This shows how the traditional formulations of 
DSR can be recovered from a single 5d action principle. We then 
discuss the DSR physics in these two bases, equations of motion, 
speed-momentum relation, boosts and length contraction. In 
section \ref{sec_mendes}, it is pointed out that another basis 
found in the literature does not fit in the gauge-fixing scheme.

In section \ref{sec_many}, we discuss the description of 
many-particle systems and the definition of a total momentum. In 
the final section, we argue that physical 
predictions should be independent of the choice of gauge fixing 
condition while the choice of particular 4d coordinates should 
correspond to the choice of specific measurement protocols for 
$x$ and $p$ in the context of general relativity providing us 
with a fluctuating space-time geometry.

We use capital letters $X$ and $P$ to denote coordinates in the 
initial phase space and small letters $x$ and $p$ to denote 
coordinates on the reduced phase space. 
%We use the following convention regarding indices: the 
Capital Latin indices refer to 5 dimensional spacetime, Greek to 
4 dimensional spacetime, while small Latin to 3 dimensional 
space. Contractions on 4d variables, such as $p^2$ and $x \cdot 
p$, are all defined using a metric $\eta_{\mu\nu} = diag(+---)$.

%%%%%%%%%%%%%%%%%%%%%%%%%%%%%%%%%%%%%%%%%%%%%%%%%%%%%%
\section{A Review of the Relativistic Particle \label{sec_review}}
%%%%%%%%%%%%%%%%%%%%%%%%%%%%%%%%%%%%%%%%%%%%%%%%%%%%%%

%%%%%%%%%%
\subsection{The Action and the Legendre transform}

Let us consider the free relativistic spinless particle of mass 
$m$ in 4 dimensions. 

In the Lagrangian formulation, the action is given by the line 
element in the flat metric $\eta=(+---)$
\be
\label{s_leg_form} 
\ss\equiv \,m\int ds= \, m\int d\tau 
\sqrt{\dot{x}^{\mu}\dot{x}_{\mu}}=\int d\tau\lll,
\ee
which we parametrize in terms of an arbitrary time parameter. The Euler-Lagrange 
equations specify the symplectic form and phase space structure. 
The canonical momentum is defined as
\be
\label{p1} 
p_{\mu}\,=\,\frac{\partial\lll}{\partial \dot{x}^{\mu}} = 
m\frac{\dot{x}_{\mu}}{\sqrt{\dot{x}^{\mu}\dot{x}_{\mu}}}
\ee 
and  the Poisson bracket on the phase space is canonical
\be 
\{x_{\mu}, p_{\nu}\}=\eta_{\mu\nu}.
\ee
Applying the Legendre transform, we check that $\lll$ is simply equal to 
$\dot{x}^{\mu}p_{\mu}$, so that the Hamiltonian vanishes. 
Nevertheless, the momentum $p_\mu$ satisfies $\hh = 0$ with
\be 
\label{hcon_sr}
\hh=p^2-m^2,
\ee
which we call the Hamiltonian  constraint. Finally the free particle action
is written in the first order formalism
\be 
\label{S_ham_form}
\ss= \int  d\tau(\dot{x}^{\mu}p_{\mu}-\lambda\hh),
\ee
where $\lambda$ is a  Lagrange multiplier. Note that while the action (\ref{s_leg_form}) cannot describe massless particles, the action (\ref{S_ham_form}) can.

The latter action (\ref{S_ham_form}) in the hamiltonian formalism can be 
taken as a starting point for the study the relativistic particle 
as an alternative to (\ref{s_leg_form}). The constraint $\hh$ 
generates a gauge symmetry whose flow defines the equations of 
motion:
\be
\label{xpflow1}
\begin{array}{rcl}
\delta x_{\mu}&=&\{x_{\mu}, \lambda\hh\} = 2\lambda p_{\mu},\\
\delta p_{\mu}&=& \{p_{\mu}, \lambda\hh\}=0.
\end{array}
\ee
Enforcing the mass-shell condition $p_{\mu}p^{\mu}=m^2$ fixes 
the value of the Lagrange multiplier
$$
\lambda= \frac{1}{2m}\sqrt{\delta x_{\mu}\delta x^{\mu}}
$$
so that the momentum is
$$
p_{\mu}= m\frac{\delta x_{\mu}}{\sqrt{\delta x_{\nu}\delta 
x^{\nu}}} 
=m\frac{\dot{x}_{\mu}}{\sqrt{\dot{x}_{\nu}\dot{x}^{\nu}}}.
$$
Inserting this expression in the action (\ref{S_ham_form}), we 
recover the original action (\ref{s_leg_form}) defined as the 
line element.

%%%%%
\subsection{Introducing Time}
%%%%%

A time variable is a phase space function that does not commute 
with the Hamiltonian (constraint). 

The time parameter that is usually chosen is $t=x_0$. As seen in 
(\ref{xpflow1}), $x_0$ does not commute with $\lambda \hh$. Then 
the equations of motion of the other variables in terms of the 
chosen time are obtained by the flow of the Hamiltonian 
constraint. Indeed, since $\{\lambda\hh, x_\mu\}=-2\lambda p_\mu\ne 
0$ and $\{\lambda\hh, p_\mu\}=0$, we have \be \label{vphys1} 
\dot{x}_i\equiv \f{\delta x_i}{\delta x_0}=\f{p_i}{p_0},\qquad 
\dot{p}_i=0. \ee The first equation defines a notion of
speed $v_i=\dot{x}_i = p_i/p_0$.

Substituting $t=x_0$ into the action (\ref{S_ham_form}) gives 
$S=\int dt\, \left[\dot{x}_ip_i-p_0-\lambda(p^2-m^2)\right]$. 
Solving the constraint, we get $p_0=\pm \sqrt{\v{p}^2+m^2}$. 
Therefore our action reduces to \be \nn S=\int dt\, 
(\dot{x}_ip_i-H),\ee where the Hamiltonian now does not vanish 
and is $H=\pm\sqrt{\v{p}^2+m^2}$, the sign depending on whether 
we consider the positive or negative energy branch.

\medskip

Another ``canonical'' choice of time variable is $t=D/m$, where 
$D\equiv x_\mu p^\mu$ is the dilatation operator. This is an 
acceptable time variable as $D$ does not commute with $\hh$,
$$ \left\{\lambda\hh,\f{D}{m}\right\}=-2\lambda\f{p^2}{m}\ne0. $$
Now we get the relativistic velocity \be \dot{x}_\mu\equiv 
m\f{\delta x_\mu}{\delta D}=m\f{p_\mu}{p^2}=\f{p_\mu}{m}. \ee 

If we want to make a time-space splitting with $D$ as the time, 
we need to identify space variables which commute with $D$ so 
that $D$ can be considered as establishing a legitimate clock 
system. The standard space coordinates $x_i$'s do not commute 
with $D$. Nevertheless, we can define normalized space variables 
$y_i\equiv (x_i\sqrt{p^2})/m$, which do commute with our chosen 
time $D/m$ and which are equal to the standard space coordinates 
on-shell. Their velocities are
$$
\dot{y}_i=\f{p_i}{\sqrt{p^2}},
$$
and reduce to $p_i/m=\gamma v_i$ on the mass-shell. However, 
these speeds are not bounded by the speed of light, contrary to 
(\ref{vphys1}), and so cannot be regarded as physical speeds. 
This shows that we can choose theoretically any time variable to 
describe the evolution of the relativistic particle, the standard 
time coordinate $x_0$ remains the physical coordinate time.

Actually, thinking of $D/m$ as the proper time along the 
trajectory of the particle, i.e as the projection of the position 
vector $x_\mu$ along the particle momentum $p_\mu$, it is natural 
to introduce the transversal position variables \be 
\tl{x}_\mu=\left(\eta_{\mu\nu}-\f{p_\mu p_\nu}{p^2}\right)x_\nu 
\ee which are orthogonal to $p_\mu$, $p_\mu \tl{x}^\mu=0$. 
Although $\tl{x}$ is defined as a 4-vector, it has only three 
(independent) components. And the  vector $(D/m, \tl{x})$ defines 
a coordinate system -the parallel and transverse projection of 
the 4-vector $x_\mu$ along the momentum $p_\mu$. The 
$\tl{x}_\mu$'s do not commute with the dilatation time $D/m$, but 
we can rescale them by the factor $\sqrt{p^2}/m$. The interesting 
feature of this choice of space coordinates is that the 
$\tl{x}_\mu$ do commute with the Hamiltonian constraint. They are 
actually Dirac observables and they do not evolve (in time): the 
gauge-fixed Hamiltonian describing the evolution of the space 
coordinates $\tl{x}$ the particle with respect of the time $D/m$ 
vanishes.

%%%%%%%%%%
%\subsection{Dirac Observables}

%Let us finish this review of the relativistic particle by 
%insisting on the fact that the position coordinates of the 
%particle do not commute with the Hamiltonian constraint and thus 
%are not Dirac observables: they are not physical quantities. 
%Therefore, we cannot operationally define the speed of the single 
%particle as the difference of coordinates since we can not 
%measure these coordinates. We need to consider at least systems 
%of two or more particle to have access to such point of view.

It is possible to generalize this remark and construct position 
operators for the relativistic particle which are Dirac 
observables. They would commute with the Hamiltonian and be 
constant of the motion. Actually, all Dirac observables of the 
relativistic particle are generated by the Poincar\'e generators 
$p_{\mu}$ and $ j_{\mu\nu}=x_{ \mu}p_{\nu}- p_{\mu}x_{\nu}.$ From 
these, one can construct position operators corresponding to the 
two previous choices of time $x_0$ and $D$:
$$
X^{(x_0)}_\mu(t)=\f{j_{\mu\nu}v^\nu+tp_\mu}{p.v}=x_\mu+\f{p_\mu}{p_0}(t-x_0),
$$
with $v=(1,0,0,0)$, and
$$
X^{(D)}_\mu(t)=\f{j_{\mu\nu}p^\nu+tp_\mu}{p^2}=x_\mu+\frac{p_\mu}{p^2}(t-D)
=\tl{x}_\mu+\f{p_\mu}{p^2}t.
$$
These are relational observables indicating the evolved positions at the time $t$ (with respect to the chosen clock $x_0$ or $D$).
A detailed analysis of the properties and quantization of these Dirac observables is under investigation \cite{FGL}.

%%%%%
\subsection{Gauge Fixing and the Dirac bracket}

The choice of a time variable can be regarded as an explicit 
gauge fixing that breaks the symmetry of the action under time 
reparametrization. After gauge fixing, the symplectic form on the 
reduced phase space is given by the Dirac bracket. Given a 
constraint $\hh$ and a gauge fixing condition $\cc$ such that 
$\{\cc,\hh\}\neq0$, the Dirac bracket is defined as
\begin{eqnarray}
\{\phi,\psi\}_D & = & \{\phi,\psi\}
-\{\phi,\cc\}\left(\frac{1}{\{\hh,\cc\}}\right)\{\hh,\psi\} \nn\\
&& -\{\phi,\hh\}\left(\f{-1}{\{\hh,\cc\}}\right)\{\cc,\psi\}.
\end{eqnarray}

For the standard time choice $\cc=x_0-t$, where $t$ is a free 
parameter, we obtain
\begin{eqnarray}
\{x_0,p_0\}_D&=&0, \nn\\
\{x_i,p_0\}_D&=& -v_i, \nn\\
\{x_\mu,x_\nu\}_D&=&0.
\end{eqnarray}
Note that $p_0$ generates the usual Hamiltonian flow under the 
Dirac bracket and is actually the true physical Hamiltonian dictating the time evolution with respect to the clock $x_0$.

The important remark linking the Dirac observables to the Dirac bracket defining the symplectic structure on the reduced phase space is the following equalities:
\begin{eqnarray}
\{x_\mu,p_\nu\}_D&=&\{X^{(x_0)}_\mu(t),p_\nu\}, \nn\\
\{x_\mu,x_\nu\}_D&=&\{X^{(x_0)}_\mu(t),X^{(x_0)}_\nu(t)\},
\end{eqnarray}
for any fixed parameter $t$. These formula summarize the explicit link between the gauge fixing procedure allowing us to generate a time evolution and the algebra of Dirac observables which are constants of motion.

In the case of the Lorentz invariant gauge fixing defined by the 
dilatation $\cc=D/m-t$, we compute
\begin{eqnarray}
\{x_{\mu},p_{\nu}\}_D&=&\eta_{\mu\nu}- \frac{p_{\mu}p_{\nu}}{m^2}, \nn\\
\{x_{\mu},x_{\nu}\}_D&=&-\frac{j_{\mu\nu}}{m^2}.
\end{eqnarray}
And once again we obtain the following equalities:
\begin{eqnarray}
\{x_\mu,p_\nu\}_D&=&\{X^{(D)}_\mu(t),p_\nu\}, \nn\\
\{x_\mu,x_\nu\}_D&=&\{X^{(D)}_\mu(t),X^{(D)}_\nu(t)\}.
\end{eqnarray}
%In particular, we have:
%$$
%\{x_0,p_0\}_D=-\gamma^2v^2, \qquad \{x_i,p_0\}_D=-\gamma v_i.
%$$
%In this case, let us point out that $p_0$ is not the hamiltonian. 

%%%%%%%%%%%%%%%%%%%%%%%%%%%
\section{Deformed Special Relativity \label{sec_dsr}}

%%%%
\subsection{A 5d Action Principle}

We now turn to the relativistic particle in the context of 
Deformed Special Relativity (DSR). The idea behind DSR is that 
the four-dimensional momentum space of the particle is curved. More precisely, we choose a constant 
positive curvature i.e. the de Sitter space. This is implemented as a constraint in the action.

The full action for the particle in the Hamiltonian formalism is
\be
\label{S5d_dsr}
S_{5d}=\int d\tau \,\left[\dot{X}_AP^A 
-\Lambda \hh_{5d} -\lambda \hh_{4d}\right].
\ee
The coordinates  $X_A$ and $P^A$ form a ten dimensional phase space with the usual 
symplectic structure $\{X_A,P_B\}=\eta_{AB}$. The first constraint
\be
\hh_{5d}=P_AP^A+ \ka^2
\ee
imposes that the  physical 4d momentum effectively lives on the four-dimensional 
de Sitter space $\SO(4,1)/\SO(3,1)$. The second constraint
\be 
\hh_{4d} = P_4-M
%\hh_{4d} = \hh_{4d}(P_4,M)
\ee
is identified as the mass-shell constraint $\hh$ in 
(\ref{hcon_sr}) and generates the Hamiltonian flow 
on the reduced phase space. The mass $M$ is a function of the rest mass $m$ and the universal mass scale $\kappa$. 
We will show in the following section how $\hh_{4d}$ becomes the Hamiltonian constraint of the deformed relativistic particle and how it is actually a $\kappa$-dependent deformation of the traditional $p^2-m^2$.
Actually, instead of $\hh_{4d} = P_4-M$, we could take $\hh_{4d} = V^AP_A-M$ for any (unit) 5d space-like vector.

We remark that the structure of this action is very similar to the reformulation of general relativity as a perturbed BF theory with a $\SO(4,1)$ gauge symmetry studied by Freidel and Starodubtsev in \cite{l&a}. Indeed, the following action  has the Einstein equations as classical equations of motion:
\bes
S_{FS}[B,A]&=&\int B^{IJ}\w F_{IJ}[A]
-\f{\beta}{2}B^{IJ}\w B_{IJ} \nn\\
&&-\f{\alpha}{4}\int V^M\epsilon_{IJKLM}B^{IJ}\w B^{KL},
\ees
where $V^M$ is a unit vector and $I,..,M=0..4$ are $\so(4,1)$ indices. The coupling constants $\alpha$ and $\beta$ are related to the cosmological constant and the Immirzi parameter (which is related to CP violation). The kinematical term of the action is the topological BF gauge theory. The second term is a $\SO(4,1)$-invariant mass term which introduces the cosmological constant. The third term explicitly breaks the $\SO(4,1)$ gauge symmetry down to a $\SO(3,1)$ gauge symmetry and allows to recover general relativity as a sector of the topological BF theory. We believe that the similarity between this reformulation of general relativity as a $\SO(4,1)$ gauge theory and the 5d action for deformed special relativity might point to the fact that our 5d action would actually describe the dynamics of (test) particles in this framework.

%{\bf [JKG] This is an interesting remark, but it is rather unclear to me what do you have in mind. Please elaborate!}

\medskip

The symmetries of the action are generated by those 
operators that commute with both of the constraints. It is 
straightforward to check that the 4 dimensional angular momentum operators,
\be 
\label{Lsymgen}
J_{\mu\nu} = X_\mu P_\nu-X_\nu P_\mu,
\ee
commute with the two constraints. The $J_{\mu\nu}$'s  together with the $P_\mu$'s are the generators of the ten-dimensional Poincar\'e Lie algebra. We also refer to these operators as the spacetime boosts $N_i = J_{0i}$ and the rotations $M_{ij} = J_{ij}$.

%%%%
\subsection{Gauge Fixing down to DSR's}

The proposed action $S_{5d}$ is defined independently of any choice of  
``basis'' or coordinate system on de Sitter. Standard theories of 
DSR are then to be obtained through a gauge fixing of the 5d
$\kappa$-shell constraint $\hh_{5d}$.

More precisely, a gauge fixing is defined as a phase space function $\cc$ which does not commute with $\hh_{5d}$:
$$
\{\hh_{5d},\cc\}\ne 0.
$$
The additional constraint $\cc=0$ turns $\hh_{5d}$ into a second class constraint and imposing $\hh_{5d}=\cc=0$ will reduce the initial 10-dimensional phase space to a more traditional 8-dimensional phase space for a relativistic particle. The symplectic structure induced on the reduced phase space is the Dirac bracket defined as previously by:
\bes
\{\phi,\psi\}_D & = & \{\phi,\psi\}
-\{\phi,\cc\}\left(\frac{1}{\{\hh,\cc\}}\right)\{\hh,\psi\} \nn\\
&& -\{\phi,\hh\}\left(\f{-1}{\{\hh,\cc\}}\right)\{\cc,\psi\} \nn.
\ees
A choice of proper coordinates on the reduced phase space are $(x_\mu,p_\mu)$ that commute with both $\hh_{5d}$ and $\cc$:
\be
\{x_\mu,\hh_{5d}\}=\{x_\mu,\cc\}=
\{p_\mu,\hh_{5d}\}=\{p_\mu,\cc\}=0.
\ee
Their Dirac bracket with any phase space function is exactly equal to their Poisson bracket with that same function:
$$
\forall F(X,P),\, \{x,F\}_D=\{x,F\},
\quad
\{p,F\}_D=\{p,F\}.
$$
We usually choose $p_\mu$ as functions of solely the 5d momentum 
coordinates $P_A$ while the $x_\mu$ are a mixture of $X_A$ and 
$P_A$. The $p_\mu$'s are a choice of coordinate system of the 
4-dimensional De Sitter space, while the $x_\mu$ are a choice of 
generators of the translations on this same De Sitter space. 
Usual DSR theories are formulated through such a choice of 
coordinates $(x_\mu,p_\mu)$, which is called a choice of basis. 
Here we identify the ambiguity as the choice of a gauge fixing 
condition.
%Here we reduce the ambiguity to the simple choice of a gauge fixing condition.

Moreover, now having ten phase space functions $\hh_{5d}, \cc, 
x_\mu, p_\mu$, we can invert the relation between the 4d 
coordinates and the 5d coordinates and express the $X_A$ and 
$P_A$ in terms of $x_\mu,p_\mu$ and $\hh_{5d},\cc$. Since 
$\hh_{5d}$ and $\cc$ commutes with the $x$'s and $p$'s, we can 
fix them at some particular values and express $X,P$ in term of 
$x,p$ and the real parameters $\ka$ and $T$. Actually $\cc$ is 
not needed to express the 5d momentum coordinates $P_A$ which are 
a function of $p_\mu$ and $\kappa$ only; $\cc$ is necessary, 
however, to express the 5d coordinates $X_A$ in terms of the 4d 
coordinates $x_\mu$. Thus, the condition $\cc$ is crucial to 
understanding the ``meaning" of the fifth space-time coordinate 
$X_4$. 

Finally, we have the interesting property that any phase space 
function $\vphi(X,P)$ commuting with both the constraint 
$\hh_{5d}$ and the gauge fixing condition $\cc$ is actually a 
function of solely the $x,p$'s. In particular, this implies that 
the $x,p$ variables form a closed algebra under the Poisson 
bracket and define a true 4d phase space without having 5d 
variables appearing in their commutation relations. This also 
implies that it should be possible to quantize the DSR particle 
as a 4d system, without necessarily having to deal with the 5d 
variables. Nevertheless, we feel that a proper quantization 
should be done at the fully five-dimensional context without 
introducing any particular gauge fixing.

%\medskip

Since the different DSR theories can be seen as particular gauge 
fixing of this 5d action, we expect that the physical predictions 
of the theory should be independent from the gauge fixing choice 
and depend solely on the details of the 5d action. Moreover, as 
we know that gauge fixing and quantization do not commute, we 
expect that the quantization of the relativistic particle in DSR 
should be done from the 5d point of view and not in a particular 
basis of a DSR theory.

%\medskip

%In the following sections, we follow the gauge fixing procedure and provide the explicit 
%gauge fixing conditions leading to the Snyder basis and the 
%bicrossproduct basis. This shows how to recover traditional formulation of DSR from this unique 5d action principle.
%We then discuss the DSR physics in these two basis, equations of motion, speed-momentum relation, boosts and length %contraction.

%We will use small letters, $x$ and $p$, to denote coordinates on the reduced phase space. 
%Also, contractions on 4d variables, such as $p^2$ and $x \cdot 
%p$, are all defined using a metric $\eta_{\mu\nu} = diag(+---).$

%%%
\subsection{The 5d interpretation of DSR}

We note that the 5d action  (\ref{S5d_dsr}) can be rewritten in a 
suggestive manner as \be S_{5d}=S_{4d}+S_4 -S_I. \ee Here 
$S_{4d}$ is the standard (undeformed) relativistic particle, \be 
\nn S_{4d}=\int d\tau\,\left[\dot{X}_\mu P^\mu -\lambda (P_\mu 
P^\mu-\tl{M}^2)\right], \ee with a renormalized mass 
$\tl{M}^2=M^2-\kappa^2$. This works if and only if $M>\ka$. $S_4$ 
is an extra degree of freedom,
$$
S_4=-\int d\tau\, \dot{X}_4 P_4,
$$
and $S_I$ is an interaction term enforcing the 
energy conservation of the coupled system of the particle and the 
extra degree of freedom,
$$
S_I= \Lambda\int d\tau\, (P_\mu P^\mu-P_4^2 +\kappa^2).
$$
The extra degree of freedom comes in with an negative energy and thus resembles a conformal mode. This suggests to consider DSR as describing a standard relativistic particle  coupled to some effective degree of freedom, which could come from some effective description of (quantum) gravity. This point of view will be investigated in more details in future work.

%%%%%%%%%%%%%%%%%%%%%%%%%%%%
\section{Snyder Basis \label{sec_snyder}}
%%%%%%%%%%%%%%%%%%%%%%%%%%%%

The gauge fixing condition $\cc$ defining the Snyder basis is the 5d dilatation
\be
\cc = \dd-T=X_AP^A-T
\ee
where $T$  is an arbitrary real parameter. In the following, we study the 
reduced phase space of the particle, after the gauge is imposed, 
as well as the equations of motion and the kinematics of the 
particle in this gauge.

\subsection{Reduced phase space}

Let us introduce the following four-dimensional position and 
momentum
\be
\begin{split}
p_\mu&\equiv \kappa \f{P_\mu}{P_4}, \\
x_\mu&\equiv \f{1}{\kappa}J_{\mu 4}=\f{1}{\kappa}(X_\mu P_4-X_4 
P_\mu).
\end{split}
\label{XPsnyder}
\ee
It is straightforward to check that these coordinates commute with both $\hh_{5d}$ and $\cc=\dd-T$.
In terms of these 4d  variables, the kinetic part of $S_{5d}$ can be written up to boundary terms as
\be 
\label{LkinSnyder}
\dot{X}_AP^A = \dot{x}_\mu p^\mu - p_\mu 
\f{x\cdot p }{\ka^2-p^2} \dot{p}^\mu,
\ee
where in addition to  the usual $\dot{x}_\mu p^\mu$ term, there is also a new term that 
depends on $\ka$. Note that to arrive at the above formula, we 
explicitly imposed the constraint $\hh_{5d}=0$.

Now, it is easy to compute the Poisson bracket of the 4d 
variables either directly from the definitions (\ref{XPsnyder}) 
and the 5d symplectic structure or from the kinematical part of 
the action given in (\ref{LkinSnyder}). The resulting structure is 
\be 
\begin{split}
\{x_{\mu}, x_{\nu}\}&= -\frac{1}{\kappa^2}J_{\mu\nu}, \\
\{x_\mu,p_\nu\}&= \eta_{\mu\nu}-\frac{p_\mu p_\nu}{\kappa^2}, 
\end{split} \label{XPSnyder_bracket} \ee
where the Lorentz generators are: 
$$
J_{\mu\nu}=(X_\mu P_\nu-X_\nu P_\mu)
=(x_\mu p_\nu-x_\nu p_\mu).
$$
This deformed symplectic structure defines the Snyder non-commutative space-time.
When $\kappa$ goes to infinity, we recover the standard phase space of the relativistic particle. 

\medskip

Starting from (\ref{LkinSnyder}), we note that the kinetic term is trivialized to $x_\mu^\prime\dot{p}^\mu$ with
the new set of $4d$ position coordinates:
\be 
x_\mu^\prime = x_\mu + p_\mu \f{x \cdot p}{\ka^2-p^2}.
\ee 
The symplectic structure expressed with these new coordinates is simply
$\{x_\mu^\prime,p_\nu\}=\eta_{\mu\nu}.$ Since the new primed 
coordinates are defined only in terms of the unprimed ones, they 
still commute with both the Hamiltonian constraint and the 
gauge-fixing condition. The natural issue is the physical meaning and interpretation of these new coordinates $x_\mu^\prime$.

\medskip

Since the 4d variables $x_\mu$ and $p_\mu$ commute with the 5d 
$\kappa$-shell constraint $\hh_{5d}$, we can 
interpret $x,p$ as Dirac observables of the 5d system (with respect to the 5d $\ka$-shell constraint, not the $P_4=M$ mass-shell constraint). On the 
other hand, this means that considered as operators they leave 
the De Sitter space invariant: the $x_\mu$'s generate the translations 
on the De Sitter space, which has become our 4d momentum space. 
Since $x_\mu,p_\mu$ also commute with the gauge fixing condition $\dd$,
we can fix $\hh_{5d}=\dd=0$ in the 
ten-dimensional phase space without interfering with the $x,p$ 
variables and reduce the phase space to eight-dimensions. By this, we mean that the gauge fixed bracket -the Dirac bracket- of $x_\mu,p_\mu$ with any function will be equal to their Poisson bracket. This 
also allows us to invert the definition \Ref{XPsnyder} and 
express the 5d coordinates $X,P$ in terms of $x,p,\kappa,T$. The 
condition $\hh_{5d}=0$ is used to obtain the 5d momentum $P_A$, 
while the condition $\dd=0$ gives the expression for the 5d 
position $X_A$. We denote the resulting functions $\tl{X}$ and 
$\tl{P}$ in order to avoid confusion with the original variables 
$X,P$,
\be 
\begin{split} \tl{P}_4 &=\f{\kappa}{\sqrt{1-\f{p^2}{\kappa^2}}}, 
\qquad\;\; \tl{P}_\mu =\f{p_\mu}{\sqrt{1-\f{p^2}{\kappa^2}}}, \\  
\tl{X}_4 &=\f{1}{\kappa\sqrt{1-\f{p^2}{\kappa^2}}}(x\cdot p -T), \\
\tl{X}_\mu &=x_\mu\sqrt{1-\f{p^2}{\kappa^2}} + 
\f{p_\mu}{\kappa^2\sqrt{1-\f{p^2}{\kappa^2}}}(x\cdot p -T),
\end{split} \label{tlXPSnyder}
\ee 
or in a more condensed format:
$$
\tl{X}_\mu=\f{\ka}{\tl{P}_4}x_\mu^\prime-T\f{\tl{P}_\mu}{\ka},
$$
$$
\tl{X}_4=\f{\tl{P}_4}{\ka^2}x.p-T\f{\tl{P}_4}{\ka}.
$$
On the one hand, this explains the origin of the $x_\mu^\prime$ coordinates: they are the four-dimensional sector of the 5d coordinates (at $T=0$). On the other hand, the inversion formula reveals the meaning of the fifth space-time coordinate (in the Snyder basis): it is the dilatation $D=x.p$ on the 4-dimensional relativistic particle.

Finally, we can compute the Dirac bracket corresponding to our 
gauge fixing. We obtain:
\be
\begin{split}
\{X_A,X_B\}_D&=\f{-1}{P_CP^C}J_{AB}=\f{1}{\kappa^2}J_{AB},\\
\{X_A,P_B\}_D&=\eta_{AB}-\f{P_AP_B}{P_CP^C}=\eta_{AB}+\f{P_AP_B}{\kappa^2}. 
\end{split}
\ee
%This algebra is very similar to the  Snyder algebra (\ref{XPSnyder_bracket}) for the 4d coordinates $(x,p)$.
First, we point out that the Dirac bracket of the 4d 
variables $x_\mu,p_\mu$ is equal to the Poisson bracket since 
they commute with $\hh_{5d}$: the bracket is unchanged and we 
remain with the commutation relations \Ref{XPSnyder_bracket}. 
Second, we have the following relations on the 5d variables:
\be 
\begin{split}
\{\tl{X}_A,\tl{X}_B\}&=\{X_A,X_B\}_D, \\
\{\tl{X}_A,\tl{P}_B\}&=\{X_A,P_B\}_D, \\
\{\tl{P}_A,\tl{P}_B\}&=\{P_A,P_B\}_D. \end{split}
\ee

\subsection{4d DSR Physics}

%Also, $x \cdot p$ is the coordinate of $x_\mu$ measured along the 
%worldline of the particle. Here, therefore, is a relationship 
%between $X_4$ and the 4d gauge fixing condition $x\cdot p-T$ 
%which we described earlier. {\bf This paragraph is confused. It 
%is also out-of-place.}

So far, we have derived the 4d phase space by gauge fixing the 5d 
phase space. Now we are interested in the dynamics of the resulting 4d 
particle. By this we mean writing down the 4d Hamiltonian 
flow, choosing a time variable and the equations of motion.

The fifth component of momentum $P_4$ is equal to 
$\sqrt{\kappa^2-P_\mu P^\mu}$ on the $\kappa$-shell and is a 4d 
Lorentz invariant. Moreover, as $P_4=\kappa/\sqrt{1-p^2/\kappa^2}$ is simply a 
function of $p^2$, we can make the 4d Hamiltonian constraint 
$\hh_{4d}=P_4-M$ of the same form as (\ref{hcon_sr}), i.e. 
$$
\hh_{4d}=p^2-m^2=0,
$$
with
$$
M=\f{\ka}{\sqrt{1-\f{m^2}{\ka^2}}}, \quad
m=\ka\sqrt{1-\f{\ka^2}{M^2}}.
$$
Note that the rest mass $m$ is bounded by the universal mass scale $\ka$, while $M$ is necessarily larger than $\ka$.
%Note that imposing this constraint greatly simplifies many expressions such as (\ref{tlXPSnyder}).
Computing  the Hamiltonian flow of the position variables gives
$$
\delta x_\mu=\{\lambda \hh, x_\mu\}=-2\lambda 
p_\mu\left(1-\f{p^2}{\kappa^2}\right),
$$
leading to the same speed formula $v_i\equiv\dot{x_i}=p_i/p_0$ as in 
special relativity if we choose $t=x_0$ as our time. Therefore, 
we have the same relation between the speed and momentum as in 
special relativity, $p_\mu=(\gamma m, \gamma m v_i)$.

\medskip

The similarity between Snyder basis physics and standard special 
relativity suggests that there should be a direct mapping between 
the two. Indeed one can introduce {\it commutative} coordinates $y_\mu, 
q_\mu $ in the Snyder basis $x_\mu, p_\mu$. Let
\be 
\begin{split} y_\mu &=\kappa 
\f{x_\mu}{\tl{P}_4}=x_\mu\sqrt{1-\f{p^2}{\kappa^2}}, \\
q_\mu & 
=\f{\tl{P}_4p_\mu}{\kappa}=\f{p_\mu}{\sqrt{1-\f{p^2}{\kappa^2}}}. 
\end{split}
\ee
Then these new coordinates satisfy trivial canonical relations
$$
\{y_\mu,y_\nu\}=0, \quad \{y_\mu,q_\nu\}=\eta_{\mu \nu}.
$$
The Hamiltonian flow on the phase space $(y,q)$ is equivalently 
generated by $p^2$ or $q^2$. Thus everything looks like we had 
rescaled the standard flat metric $\eta_{\mu\nu}$ of the 
Minkowski by a momentum-dependent factor $(1-p^2/\kappa^2)$.  
This type of momentum-dependent metric naturally appearing in the Snyder basis allow an explicit link with the rainbow metric formalism advocated in \cite{rainbow}.
When its rest mass $m$ reaches the maximal bound $\ka$, the metric $dx^2=dy^2/(1-p^2/\ka^2)$ goes to infinity. In some sense, when $m=\ka$, the particle freezes -it doesn't move.
Finally the gauge fixed  5d variables have a simple expression in terms of $y$ and $q$:
$$
\tl{P}_\mu=\pi_\mu,\quad 
\tl{X}_\mu=y_\mu+\f{q_\mu}{\kappa^2}(y.q-T).
$$

\medskip

The Dirac observables for the DSR particle are simply $p_\mu$ (or 
the rescaled $q_\mu$) and $J_{\mu\nu}$, which generates the usual 
(undeformed) Poincar\'e algebra. One can construct position 
Dirac observables the same way as done for the undeformed 
relativistic particle. Indeed, despite the deformed symplectic 
structure, the position operators $x^{(D)}_\mu=x_\mu+p_\mu/p^2(T 
-x\cdot p)$, with an arbitrary parameter $T$, still commute with 
the Hamiltonian constraint $p^2$ and are therefore Dirac observables.

From the 5d perspective, it is even more straightforward to 
obtain such position Dirac observables commuting with both constraints, $\hh_{5d}\sim P_AP^A$ and $\hh_{4d}\sim P_4$. Indeed, the positions $X_\mu$ commute with $P_4$. Nevertheless these 
variables do not commute with the $\kappa$-shell condition 
$P_AP^A=-\ka^2$. We can still use the same construction as for the 4d relativistic particle applied to our 5d particle and write down position Dirac observables on 
De Sitter as $X_\mu+P_\mu/P^2(T -X.P)$. These actually are the 
previous $x^{(D)}_\mu$ re-scaled by $\kappa/P_4$.

\subsection{Relativistic effects?}

$L^2\equiv x_\mu x^\mu$ is the natural Lorentz invariant quadratic function of the Snyder basis positions $x_\mu$ and will be therefore considered as defining the Minkowski metric in our 
non-commutative space-time. In terms of the 5d variables, it 
reads as:
\bes
\kappa^2\,L^2 &=& J_{4\mu}J^{4\mu} \nn\\
&=& P_4^2X_\mu X^\mu + X_4^2P_\mu P^\mu -2X_4P_4X_\mu P^\mu.
\ees 
In the Snyder basis, the metric is unchanged, the action of the Lorentz transformations on $(x,p)$ is as unmodified and the speed-momentum relation is the same as in special relativity, therefore we do not expect any DSR effect on the motion of the single DSR relativistic particle. The spectrum of the $L^2$ at the quantum level can be found in \cite{DL}.

We expect DSR effects to appear for multi-particle systems through a modification of the energy-momentum conservation and deformed scattering amplitudes. This should be connected to the fact that there does not exist any non-trivial deformation of the Poincar\'e symmetry as a Lie algebra, but there does exist non-trivial deformations as a Hopf algebra structure: $\kappa$-deformed Poincar\'e can always be re-written as the standard Poincar\'e Lie algebra while all the information about the deformation will be contained in the co-product i.e the laws of addition and conservation of energy-momentum.

\medskip

A last comment is that if we allow position-momentum mixing in the length operator and therefore momentum-dependent metrics, we have access to Lorentz invariants other than $x_\mu x^\mu$, such as the rest mass $p^2$ and the dilatation $x.p$. For instance, the 5d invariant metric has an interesting 
expression in terms of the 4d variables:
\bes 
\tl{X}_A\tl{X}^A&=&
x_\mu x^\mu\left(1-\f{p^2}{\kappa^2}\right)+\f{T(x.p-T)}{\kappa^2}, \nn\\
&=&y_\mu y^\mu+\f{T(y.q-T)}{\kappa^2}.
\ees
On the 5d light-cone, $T=X_AP^A=0$, this metric reduces to the rescaled flat metric which is thus the natural Lorentz invariant metric from the 5d perspective. Nevertheless, the physical interpretation of this metric at $T\ne 0$ is not yet clear to us. For traditional 4d physics, $T$ does not evolve during the free motion of the system. This could nevertheless be affected by the introduction of a non-trivial dynamics (through forces).
See however the discussion of the metric in the bicrossproduct basis under equation (\ref{NXbcp}) below.

%Using  simple representations of $\SO(4,1)$. The 5d dilatation is related to the Casimir of the representation:
%\be
%\f{1}{2}J_{AB}J^{AB}=
%\Delta(\pi_A\pi^A)-\dd(\dd+d+2).
%\ee
%{\bf Check signs because i did the calculations only in the Euclidean case!}

%%%%%%%%%%%%%%%%%%%%%%%%%%%%%%%%%%%%%%%%%%%%%%%%%%%%%%%%%%%%%%%%%%%%%%
\section{Bicrossproduct Basis \label{sec_bcp}}

To obtain the bicrossproduct version of DSR, we choose the gauge 
fixing condition $\cc$ as
\be
\cc =\frac{X_0-X_4}{P_0-P_4}-T,
\ee
where $T$ is again a free parameter. Therefore the bicrossproduct basis is a light cone gauge for the 5d action.

As in the previous section, we first study the reduced 
phase space obtained by implementing this constraint, and then 
look at various features of the resulting 4d physics. 

\subsection{Reduced phase space}

A set of 4d momentum variables that commute with both $\hh_{5d}$ and $\cc$ are
\be 
\label{bxp_pdefs}
p_0 \equiv \ka \ln \frac{P_4-P_0}{\ka}, \qquad 
p_i \equiv \frac{\ka P_i}{P_0-P_4}.
\ee
As for the position variables, we choose
\be \label{bxp_xdefs} x_0 \equiv 
\frac{1}{\ka}J_{40}, \qquad x_i \equiv 
\frac{1}{\ka}(J_{i0}-J_{i4}).
\ee 
The ten variables $(\hh_{5d}, \cc, x_\mu, p_\mu)$ parametrize the ten-dimensional phase space.
Actually, we are restricted to the sector $P_4>P_0$. If we want to parametrize the whole space, we should allow sign changes in the definition of the 4d momentum $p_\mu$.

From here, we repeat calculations similar to those presented in 
the previous section on the Snyder gauge-fixing. When we impose $P_AP^A+\kappa^2=0$, the kinetic part 
of the 5d action reduces to
\be
\label{Lkinbcp}
\dot{X}_A P^A = p_\mu \, \dot x_\mu + p_i\, x_i\, \dot p_0.
\ee
This provides the 4d action principle describing the DSR particle in the bicrossproduct basis.
This kinetic term directly gives the $\ka$-Minkowski symplectic 
structure on the 4d phase space, \be\label{sympbicross} 
\begin{split}
&\{x_0,p_0\}=1, \ \ \{x_i, p_j\}= -\delta_{ij}, \\
&\{x_0, x_i\}= +\frac{1}{\kappa}x_i, \ \ \{x_0, p_i\}= 
-\frac{1}{\kappa}p_i,
\end{split} \ee
with all other brackets vanishing. 

A similar 4d action was proposed in \cite{ghosh} as describing a particle propagating in the $AdS$ space-time. The 4d Lagragian was defined in term of $(x_\mu, \dot{x}^\mu)$ with a $AdS$ metric and the Dirac brackets taking into account the 2nd class constraints were shown to reproduce the $\ka$-Minkowski brackets. That action realizes the inverse Legendre transform of our Lagragian expressed in the variables $(x_\mu,p^\mu)$.

Solving the 5d Hamiltonian constraint $\hh_{5d}=0$ and fixing the 
gauge at $\cc=0$ allows to invert (\ref{bxp_pdefs}) and  (\ref{bxp_xdefs}) to obtain
\be 
\begin{split} \tl{P}_0 &= -\ka \sinh \f{p_0}{\ka} - 
\f{\overrightarrow{p}^2}{2\ka}e^{\f{p_0}{\ka}},
\\ \tl{P}_i &= -p_i e^{\f{p_0}{\ka}}, \\ \tl{P}_4 &= \ka \cosh \f{p_0}{\ka} - 
\frac{\overrightarrow{p}^2}{2\ka}e^{\f{p_0}{\ka}}, \\
\tl{X}_\mu  &= \ka\frac{x_\mu}{P_0-P_4} +T P_\mu, \\
\tl{X}_4  &= \ka\frac{x_0}{P_0-P_4} +T P_4.
\end{split}
\ee
The first set of relations are just the familiar definitions of the bicrossproduct 
basis in terms of planar coordinates on de Sitter.  

The Dirac bracket induced by the gauge fixing condition acts on 
the 5d variables as \be \begin{split}  \{ X_0, X_4\}_D 
&= \f{1}{2}\f{X_0-X_4}{P_0-P_4}, \\ \{X_i, P_j\}_D &= \delta_{ij}, \\
 \{ X_0, P_0\}_D &= -1 - \f{1}{2}\f{P_0}{P_0-P_4}, \\ \{ X_4, 
P_4\}_D &= +1 - \f{1}{2}\f{P_4}{P_0-P_4}, \\ \{ X_0, P_4\}_D &= 
-\f{1}{2}\f{P_0}{P_0-P_4}, \\ \{X_4, P_0\}_D &= 
-\f{1}{2}\f{P_4}{P_0-P_4}.
\end{split} \ee
As earlier, these important facts about the Dirac bracket are that, first, the Dirac bracket of $(x,p)$ with any other phase space function is exactly equal to the Poisson bracket and, second, the Poisson brackets of the $(\tl{X},\tl{P})$ variables is equal to the Dirac bracket of the $(X,P)$ variables computed above.

As in the Snyder case, we can find an alternate system of 
position variables that trivializes the $4d$ symplectic 
structure. For the bicrossproduct basis, we achieve this by 
defining
\be x_0^\prime = x_0 - \f{1}\ka x_ip^i, \qquad \text{and} \qquad 
x_i^\prime = x_i.
\ee
This choice of space-time coordinates is commutative and the couple $(x',p)$ have canonical Poisson brackets. In terms of the 5d coordinates, $x_0^\prime$ is given by
\bes
\ka\,x_0^\prime &=& J_{40} -  \frac{P_i}{P_0-P_4}(J_{i0}-J_{i4}) \nn\\
&=& (X_4P_0-X_0P_4)-X_iP_i+P_iP_i\f{X_0-X_4}{P_0-P_4}\nn.
\ees
This variable is singular at $P_0-P_4=0$. In terms of the 4d momentum coordinates, this singularity happens for infinite ``energy", $p_0 = \infty$. The $P_0-P_4=0$ is also the singularity for the bicrossproduct momentum variables (\ref{bxp_pdefs}).
Therefore one could decide that the coordinates $x^\prime$ are better mathematical space-time coordinates and that they should be the physical space-time coordinates since they do commute. However, as we will see below, the speed of a DSR particle in the $x^\prime$ basis is not bounded by the speed of light $c$ while the speed computed in the $x$ basis will be bounded by $c$ and behave appropriately. 

%%%%%%%
\subsection{4d physics}
%%%%%%%

We now turn to the physics and dynamics of the DSR particle in the bicrossproduct basis.

The 4d Hamiltonian constraint $\hh_{4d}=P_4-M$ can be re-written as:
\be
\hh_{4d}=\f{1}{2\ka}\hh +\ka -M,
\ee
$$
\hh=(2\kappa\sinh\frac{p_0}{2\kappa})^2-\overrightarrow{p}^2e^{\frac{p_0}{\kappa}},
$$
where $\hh$ goes to $p^2$ in the classical limit $\ka\arr\infty$.
$\hh$ is the Casimir of the $\ka$-deformed Poincar\'e group in the bicrossproduct basis \cite{ruegg}.

$\hh_{4d}=0$ reduces to the  mass-shell condition $\hh= m^2$ for the rest mass:
\be
m^2=2\ka (M-\ka),
\ee
where $M$ is restricted to be larger than $\ka$.

The mass-shell condition relates the space moment $\v{p}^2$ to 
the energy $p_0$ as:
$$
\v{p}^2=e^{-\f{p_0}{\kappa}}
\left(2\kappa^2\cosh\f{p_0}{\kappa}-2\ka^2-m^2\right).
$$
At zero speed, $\v{p}=0$, it is possible to invert the $\cosh$ and, assuming that $p_0$ is the (measured) energy $E$, we obtain corrections to the $E=mc^2$ formula:
\bes
p_0&=&2\kappa\ln\left(
\f{m}{2\ka}+\sqrt{1+\f{m^2}{4\ka^2}}
\right), \nn\\
&=&\kappa\ln\left(
1+\f{m}{\ka}\sqrt{1+\f{m^2}{4\ka^2}}+\f{m^2}{2\ka^2}
\right), \\
&\sim&
m-\f{1}{24}\f{m^3}{\ka^2}+\dots
\nn
\ees
A measurable DSR effect would then be an energy default with the rest energy being actually smaller than the mass $m$. Nevertheless, this relies on two assumptions. The first one is that $p_0$ is effectively the energy that we measure. More generally, we should carefully define the energy-momentum $p_\mu$ operationally as measurements. As an example, in 3d quantum gravity, the momentum is defined as a measurement of the geometry through the holonomy around the particle \cite{Freidel:2005bb}. The second issue is that $m$ might not be the true mass, but simply the bare gravitational mass. Then the mass defined above could be the renormalized mass taking into account the gravitational self-energy i.e the one that we do measure in a real experiment. This phenomenon happens in 3d quantum gravity where the bare mass $m$ (creating the conical singularity defining the particle) gets renormalized to $\ka\sin(m/\ka)$ due to gravitational effect \cite{Freidel:2005bb}. Such effects are also expected in 4d gravity \cite{york}.

The momentum light cone is defined as the limit case of a massless 
particle $m=0$. In that case, the dispersion relation simplifies 
to:
\be
|\v{p}|=\kappa(1-e^{-\f{p_0}{\kappa}}), \qquad 
p_0=-\kappa\ln\left(1-\f{|\v{p}|}{\kappa}\right).
\ee
For an arbitrary mass, it is still possible to extract the energy in 
term of the spatial momentum. For computation purpose, it is 
convenient to introduce $\alpha\equiv \exp(p_0/\kappa)$ or 
equivalently $p_0=\kappa\ln\alpha$. Then the dispersion relation 
is a second degree polynomial in $\alpha$ and can be solved 
explicitly to: \be 
\alpha_\pm=\f{(m^2+2\kappa^2)\pm\sqrt{\Delta}}{2(\kappa^2-\v{p}^2)}, 
\label{massshell} \ee
$$
\Delta\equiv m^4+4m^2\kappa^2+4\v{p}^2\kappa^2.
$$
%\bes
%\Delta&\equiv& (m^2+2\kappa^2)^2+4\kappa^2(\v{p}^2-\kappa^2)\nn\\
%&=&m^4+4m^2\kappa^2+4\v{p}^2\kappa^2. \nn
%\ees
It is straightforward to take the classical limit $\kappa\arr+\infty$ 
where we are supposed to recover undeformed special relativity:
$$
\alpha_\pm=e^{\f{p_0}{\kappa}}\sim 
1+\f{p_0}{\kappa}\sim1\pm\f{1}{\kappa}\sqrt{\v{p}^2+m^2},
$$
which allows us to identify $\alpha_\pm$ as the positive and 
negative energy modes. Let us underline that the negative energy 
mode is not given by $p_0\arr -p_0$. This is reminiscent of the 
deformation of the antipode in the quantum group language.

The equations of motions arising from the flow with 
respect to $\lambda \hh$ are
\be
\label{Eqmot_bcp}
\begin{split} \delta{x_0} &= -2\lambda(\ka\sinh \f{p_0}{\ka} + 
\frac{\overrightarrow{p}^2}{2\ka}e^{\f{p_0}{\ka}}), 
\\ \delta{x}_i &= -2\lambda p_i e^{\f{p_0}{\ka}},
\\ \delta{p_0} &= \delta{p_i} = 0.
\end{split}
\ee If we choose  $t=x_0$ as the time variable, then the speed is 
defined as usual as $v_i \equiv \delta x_i/\delta x_0.$ In this 
case, we can compute the norm of $v_i$ in terms of $p_0$, \be 
v^2=\f{e^{2\zeta}-ue^\zeta+1}{e^{2\zeta}-ue^\zeta+\left(\f{u}{2}\right)^2}, 
\label{dsrspeed} \ee which we express in term of the  
dimensionless energy $\zeta\equiv p_0/\kappa$ and dimensionless 
mass $u\equiv m^2/\kappa^2+2$. It is clear that the speed is 
still bounded by the speed of light $c=1$ (more exactly, the 
speed of massless particles). Moreover as in standard physics, 
the speed get reach $c$ if and only if $u=2$ i.e the mass 
vanishes $m=0$. Such a relation $v(p_0)$ could be checked 
experimentally, assuming than $p_0$ is the energy.

To complete the analysis, it is interesting to invert the  
relation between the speed and the momentum in order to get the 
deformed equivalent of the standard speed-momentum relation $p_\mu=(\gamma m, 
\gamma m\v{v})$. First we get the energy: \be \alpha_\pm(v)=1\pm 
\gamma 
\sqrt{\f{m^2}{\kappa^2}+\f{m^4}{4\kappa^4}}+\f{m^2}{2\kappa^2}, 
\ee where we keep the standard definition $\gamma\equiv 
1/\sqrt{1-v^2}$. In the classical limit $\kappa\arr+\infty$, we 
recover $p_0\sim\pm\gamma m$. Then we can compute the spatial 
moment as: \be \v{p}=\v{v}\left(\kappa-\f{m^2+2\kappa^2}{2\kappa} 
e^{-\f{p_0}{\kappa}}\right) 
=\v{v}\left(\kappa-\f{m^2+2\kappa^2}{2\kappa\alpha} \right). \ee
At zero speed $\v{v}=0$, the momentum $\v{p}$ vanishes as expected. But the proportionality coefficient is deformed, it is not simply the mass $m$ but it has a more complicated form:
\be
\v{p}\,\underset{v\arr 0}{\sim}\,
\v{v}\,\f{m\sqrt{1+\f{m^2}{4\ka^2}}}{1+\f{m^2}{2\ka^2}+\f{m}{\ka}\sqrt{1+\f{m^2}{4\ka^2}}}.
\ee
On the other side, in the limit $v\arr c$ where the speed reaches the speed of light, the momentum does not diverge as in special relativity and we find that $\v{p}=\kappa \v{v}$ and we have a finite effective mass $|\v{p}|/|\v{v}|$.

We conclude this analysis with the remark that if we had chosen $t=x^\prime_0$ as time, then we could have done the exact same calculations but we would have found that the speed with respect to $x^\prime_0$ would increase exponentially with $p_0$ and diverge to infinity. This is to be compared to the use of the dilatation time $D/m$ in special relativity. Therefore, we discard it as a physical time coordinate since we want to keep a maximal speed bound defined by the speed of light $c$.

\medskip

The symmetry generators (\ref{Lsymgen}), in terms of $p_\mu$ and 
$x_\mu$ coordinates, are
\be
\begin{split}
M_{ij} &= J_{ij}=x_i p_j - x_j p_i, \\
N_i &= J_{i0}= x_i\left( \ka e^{-\f{p_0}{\ka}}\sinh \f{p_0}{\ka} + \frac{\v{p}^2}{2\ka}\right) - x_0 p_i.
\end{split}
\ee
The rotations are undeformed, and the boosts are deformed in the first term.
Nevertheless, they still form an undeformed Lorentz algebra (for the deformed symplectic structure). However, their action on the space-time coordinates will be deformed. More precisely, we compute their action on the position variables $x_\mu$:
\be \label{NXbcp} 
\begin{split} \{N_i, x_0\} &= x_i - \f{1}{\ka}N_i, 
\\ \{N_i, x_j\} &= x_0\, \delta_{ij} - \f{1}{\ka}M_{ij}, \\ \left\{ 
M_{ij},x_0\right\} &= 0 
\\ \left\{ M_{ij}, x_k \right\} &= x_j \,\delta_{ik} - 
x_i\delta_{jk}.
\end{split}
\ee
The action of these generators on the momentum variables gives the well known bicrossproduct 
relations. Overall, the rotation algebra does not seem deformed 
whatsoever. On the other hand, the boost sector is different than 
usual; we therefore study this sector in more details below and compute the effect of this boost deformation on the traditional length contraction.

We conclude this subsection with a discussion of the Lorentz invariant quadratic form, the "metric", in the bicrossproduct spacetime.
In this basis, $x_\mu x^\mu$ is not a Lorentz 
invariant anymore. However we can deform it into
\be
\label{L2bcp}
L^2 = x_0^2 - \left(x_i-\f{1}{\ka}N_i\right)^2,
\ee
which can be thought of in principle as a new metric. This expression is exactly the same algebraic quantity as the Snyder basis metric, $J_{4\mu}J^{4\mu}$ in term of the 5d operators. However, since the quantity $N_i$ depends on the particle momentum, we obtain a momentum-dependent quadratic form, explicitly:
$$
L^2= x_0^2\left(1-\f{\v{p}^2}{\ka^2}\right)-x_ix_i f(p)^2-\f{2x_0}{\ka}\v{x}.\v{p}f(p),
$$
$$
f(p)=1-e^{-\f{p_0}{\ka}}\sinh\f{p_0}{\ka}-\f{\v{p}^2}{2\ka^2}.
$$
It is tempting to interpret the quadratic form (\ref{L2bcp}) as a space time metric\footnotemark.

\footnotetext{Another natural quadratic form can be derived from the 5d metric:
\be
\tl{X}_A\tl{X}^A=T\ka x^\prime_0-x_ix_ie^{-2\f{p_0}{\ka}}-T^2\ka^2.
\ee
However, the physical meaning of $T$ is not clear and the behavior of $\tl{X}_A\tl{X}^A$ under boosts is not simple.
}

The momentum-dependence in the metric is strange from the point 
of view of special relativity, but it would be much more natural 
from the general relativity perspective, where we expect the 
particle to deform the geometry of the surrounding space-time 
depending on its energy-momentum tensor. This effect may still be 
present in the DSR limit of gravity. 

This would mean that the spacetime metric would vary for 
different particles which is rather hard to accept physically. In 
fact, the metric usually has a clear operational meaning, being a 
measure of distances between spacetime points. What one should 
presumably do is to try to construct the metric operationally, in 
terms of physical rods and clocks, using the fact that, as we 
said above, one has in disposal a universal observer-independent 
scale of velocity. It is crucial that all inertial observers 
could agree on a way the metric is constructed, and that means 
exist to synchronize clocks and rods of observers at different 
points and/or moving with respect to each other with constant 
speed. We believe that the kinematical description of the 
particle presented here is a good starting point for such a 
construction. 

\subsection{Relativistic effects}

We integrate explicitly the action of the boosts on space-time coordinate and use this computation to study the time dilatation and length contraction in the bicrossproduct basis.

Consider for simplicity the action of boosts along the $i=1$
direction. Defining $\{N_1, F\} = \pa_\xi F$ for any function $F$, 
we obtain from the commutators (\ref{NXbcp}) the differential 
equations \be \label{odeX} \begin{split} \pa_\xi x_0(\xi) &= 
x(\xi) - N_1(\xi), \\ \pa_\xi x_1(\xi) &= x_0(\xi), \\ \pa_\xi 
x_2(\xi) &= -M_{3}(\xi), \\ \pa_\xi x_3(\xi) &= + M_{2}(\xi). 
\end{split} \ee To solve this system of equations, it is worth 
simplifying the right hand sides. Since $\{N_1,N_1\}=0$, the object 
$N_1(\xi)$ in the first equation above is actually a constant 
independent of $\xi$. Therefore, $N_1(\xi)$ can be evaluated once 
at a certain time in one particular reference frame. Using the 
definition for $N_1$, one has \be n_1 = \left[ 
x_1\left(e^{-p_0}\sinh p_0+ \frac{p^2}{2}\right) -x_0 p_1 
\right]_{0} \ee where the subscript $0$ denotes that the 
expression is evaluated a set of initial conditions for 
$x_\mu,p_\mu$.

Another trick is to find the functions $M_3(\xi)$ and $M_2(\xi)$. 
This can be done by considering the set of differential equations 
\beq \begin{split} \pa_\xi M_2(\xi) &= N_3(\xi) \\ \pa_\xi 
N_3(\xi) &= M_2(\xi) \\ \pa_\xi M_3(\xi) &= -N_2(\xi) \\ \pa_\xi 
N_2(\xi) &= -M_3(\xi), \end{split} \eeq whose solutions are \beq 
\begin{split} M_2(\xi) &= m_2 \cosh \xi + n_3\sinh \xi
\\ M_3(\xi) &= m_3 \cosh\xi - n_2\sinh \xi \\ N_2(\xi) &= n_2 \cosh \xi - m_3\sinh\xi \\ N_3(\xi) &= n_3\cosh\xi + m_2\sinh\xi.  
\end{split} \eeq The coefficients all denote initial values. 

These observations makes it straight-forward to write down the 
solutions to the system of differential equations (\ref{odeX}): 
\be \begin{split} x_0(\xi) &= x_0(0) \cosh \xi + x_1(0) \sinh \xi 
- n_1 \sinh\xi, 
\\ x_1(\xi) &= x_0(0) \sinh \xi + x_1(0) \cosh \xi - n_1 \cosh \xi + n_1 
\\ x_2(\xi) &= x_2(0) -n_2 + n_2 \cosh\xi - m_3\sinh\xi
\\ x_3(\xi) &= x_3(0) - n_3 + n_3 \cosh\xi + m_2\sinh\xi. \end{split} \ee

Given the equations of motion (\ref{Eqmot_bcp}), we can also 
study relativistic effects. The first of these is to check the 
dependence of the velocity $v_i = \delta x_i /\delta x_0$ on the 
boost parameter. Consider a particle that is initially at rest, 
i.e. its initial momentum $p_i=0$. Then, using the transformation 
laws of momenta $p_\mu$ \cite{rbgacjkg}, we find that \be v = 
\tanh{\xi} \eeq just like in special relativity. The result does 
not depend on the initial value of the particle's energy, thus we 
again find that all particles approach the speed of light $v=c=1$ 
after repeated boosts.

Other relativistic effects that are worth investigating are time 
dilation and length contraction. We follow the approach developed 
in \cite{Kowalski-Glikman:2001px} and consider a system of two 
particles. One particle, labeled $A$, is placed at $X^A=0$ and 
has zero momentum $P^A=0$ in some reference frame. The other 
particle, labeled $B$, is placed at $X^B=\ell$ and is also 
stationary $P^B=0$ in that reference frame. The equations of 
motion for these particles, in terms of their affine parameters 
$s_A$ and $s_B$, are \beq \begin{split} X_{0}^A(s_A,\xi) & = s_A 
\sinh p_{0}^A \cosh \xi \\ X_1^A(s_A,\xi) & = s_A \sinh p_{0}^A 
\sinh \xi 
\\ X_2^A(s_A,\xi) & = 0\\ X_3^A(s_A,\xi) & = 0 \end{split} \eeq and \beq \begin{split} X_0^B(s_B,\xi) &= 
s_B \sinh p_{0}^B \cosh \xi + \ell \sinh \xi - n_1^B \sinh \xi. \\
X_1^B(s_B,\xi) & = s_B \sinh p_0^B \sinh \xi + \ell_1 \cosh \xi - 
n_1^B \cosh \xi + n_1^B \\ X_2^B(s_B,\xi) & = \ell_2 -n_2 + 
n_2\cosh\xi 
\\ X_3^B(s_B,\xi) & = \ell_3 -n_3 +n_3 \cosh\xi. 
\end{split} \eeq

In one reference frame characterized by $\xi=0$, we define `time' 
along each particle's worldline by \be \nn \tau = 
x_0(s^\prime,0)- X_0(s,0) \ee as the difference of $x_0$ 
functions at different value of the affine parameter $s$. We can 
find how these quantities transform under boosts by introducing 
another frame with general $\xi$. Then time intervals becomes 
\beq \tau^\prime = x_0(s^\prime,\xi)- x_0(s,\xi) = \tau \cosh \xi 
\eeq for each particle. This is time dilation in its standard 
(special relativistic) form. 

We now discuss distance measurements. A naive way to define 
distance is to consider the coordinate difference \be \nn d_1 = 
x_1^A(0,0) - x_1^B(0,0) = \ell_1; \ee similarly expression in the 
other directions give $d_2$ and $d_3$. The distance interval 
between the particles in the second reference frame is computed 
by first setting $x_0^A(s_A,\xi)=x_0^B(s_B,\xi)$ to obtain a 
relation between $s_A$ and $s_B$, and then doing the subtraction 
of coordinates. The result along the direction of the boost is 
\beqa d_1^\prime &=& x_1^A(s_A,\xi) - x_1^B(s_B,\xi) \nn 
\\ &=& \frac{1}{\cosh \xi} \ell_1 + n_1^B \left( \frac{\cosh \xi 
-1}{\cosh\xi}\right). \eeqa The first term gives the usual length 
contraction effect, whereas the second term is a correction due 
to the momentum space curvature. Its significance grows with the 
original separation $\ell$ and with increasing $\xi$ values. 
Since $n_1^B$ is also a function of $p_0^B$, the correction is 
also dependant on what kind of particle is being discussed. In 
the orthogonal directions, the distances are just 
$d_2=x_2^B(s_B,\xi)$ and $d_3=x_3^B(s_B,\xi)$. In contrast to 
$d_1$, these distances become larger as the boost parameter is 
increased. 

An alternative approach for looking at length contraction is to 
define distances in terms of the combination $x_i-N_i$, as 
suggested by the invariant interval (\ref{L2bcp}). This amounts 
to a shift of the coordinates by a constant. Then, we have \be \nn
d_i = (x_i^A(0,0)-n_i^A) - (x_i^B(0,0)-n_i^B) = \ell_i - n_i^B. 
\ee In a boosted frame, the new measured distance along the 
direction of the boost is \beq d_1^\prime = \f{1}{\cosh \xi} d_1, 
\eeq the usual length contraction. In the other directions, 
however, there is an effect: \be d_2^{\prime} = d_2 - 
n_2^B(1-\cosh \xi ). \ee
Directions orthogonal to the boost are therefore
also slightly contracted. 

%%%%%%%%%%%%%%%%
\section{Mendes' Basis \label{sec_mendes}}
%%%%%%%%%%%%%%%%

Another basis was introduced by Vilela Mendes \cite{mendes} in 
the context of the requirement of stability of the symmetry Lie 
algebra underlying physical systems. Concluding that the standard 
phase space structure underlying relativistic quantum mechanics 
(based on the Poincar\'e algebra) is not stable, he introduced a 
stable deformation, which can be represented in the extended 
ten-dimensional phase space.

The Mendes choice of 4d coordinates is close to the Snyder basis 
but slightly different: \be \label{Mendesbasis} p_\mu=P_\mu, 
\quad x_\mu =\f{1}{\ka}J_{\mu 4}. \ee The 5d kinetic term then 
reads \be -\dot{X}_A P^A= \frac{1}{P_4} \left(x_\mu 
\dot{p}^\mu\right) %+ \frac{1}{2}P_AP^A \pa_\tau \left( 
%\frac{X_4}{P_4}\right).
%\int d\tau \dot{X}_A P^A=
%-\int d\tau \, \frac{1}{P_4} \left(x_\mu 
%\dot{p}^\mu\right) + \frac{1}{2}P_AP^A \pa_\tau 
%\left( \frac{X_4}{P_4}\right)
\ee which is like the standard kinetic term rescaled by the 
momentum $P_4$. 
%The second term becomes is a simple boundary term 
%(total derivative) once the constraint $\hh_{5d}=P_AP^A +\ka^2=0$ 
%is imposed. 
The Poisson brackets give the following commutators: 
\be \label{Mendesbrackets}
\begin{split} \{ x_\mu, x_\nu \} &= 
-\f{1}{\ka^2}J_{\mu\nu} 
\\  \{ x_\mu, p_\nu \} &= \eta_{\mu\nu} \f{P_4}{\ka}, \qquad \{ x_\mu, P_4 \} = \f{1}{\ka} p_\mu. 
\end{split}
\ee
An important feature of these relations is that the scaling factor $P_4$ 
appears explicitly on the right hand side and must be treated as 
a new element of the algebra. In other words, the algebra of the 
variables $x_\mu$ and $p_\mu$ in this basis is not closed, so 
that the space spanned by these eight variables cannot be 
considered as the phase space of a 4d spacetime independent of the fifth
dimension.

Mathematically, the fact that the $x,p$ algebra is not closed 
means that there does not exist any gauge fixing condition 
leading to this choice of coordinates. Indeed, if we consider a 
phase space function $\cc$ such that $\{\cc,x\}=\{\cc,p\}=0$, 
then we have $\{\cc,P_\mu\}=\{\cc,P_4\}=0$ since $\{x,p\}\sim 
P_4$. Thus $\cc$ must commute with $\hh_{5d}$ and cannot be 
considered as a gauge fixing condition. Furthermore, since $\cc$ 
must commute with $x_\mu=J_{\mu 4}$, $\cc$ must be a 
$\SO(4,1)$-invariant function of only the $P_A$'s. Thus it must 
be a function of $P_AP^A$.

We conclude that although Mendes' basis (\ref{Mendesbasis}) 
amounts to a specific choice of parametrization of the 
ten-dimensional phase space, it does not allow a straightforward 
interpretation as a 4d (non-commutative) spacetime through a 
gauge fixing of the 5d $\ka$-shell constraint $\hh_{5d}$. 
Therefore, Mendes' basis is simply still  a choice of variables 
for the same DSR theory, it does not allow the same 
interpretation as the Snyder basis or the bicrossproduct basis.

The other interesting feature of Mendes' work \cite{mendes} is 
his prescription for cross-sections. He discusses a change of 
measure in momentum space:
$$
d^3 p \,\arr\,
\f{d^3p}{\sqrt{1-\f{p^2}{\ka^2}}}.
$$
On one hand, it is exactly the change of measure identified for 
3d DSR \cite{Freidel:2005bb} when considered as an effective 
theory for quantum gravity. On the other hand, it must be 
compared to the measure derived by Snyder for 4d DSR 
\cite{snyder}:
$$
d^4p\,\arr\, \f{\ka^2d^4p}{\ka^2-p_\alpha p^\alpha}.
$$
Furthermore, Mendes postulates a trivial addition of momentum, 
$p^{tot}_\mu=p^{(1)}_\mu+p^{(2)}_\mu$. Since $p_\mu=P_\mu$, this 
fits with the proposal \cite{GL,GL2} that the five-momentum $P_A$ 
is additive, but contradicts the standard view that the addition 
of 4d momenta $p_\mu$ should be deformed.

\section{Towards many-particle states \label{sec_many}}

In this paper, we have studied the dynamics of a single free 
particle in DSR. A necessary step towards building a quantum 
field theory in a DSR space-time is to study the classical 
mechanics of many particles, in particular two particles and the 
mechanics of the center of mass frame. A natural remark is that 
if we start from the 5d action principle for the DSR particle, 
$\ka$ does not need to be a universal mass scale but could be 
assumed to be a property of the considered particle on the same 
level than the rest mass $m$.

An important issue is the definitions of the  addition of 
(energy)-momentum and of the total momentum in the center of mass 
frame. Usually, in DSR theories, one chooses a particular basis 
and writes down a deformed addition $p_\mu \oplus q_\mu \ne p_\mu 
+ q_\mu$ such that the original momenta $p,q$ and the total 
momentum $p\oplus q$ are all on the same de Sitter space for the 
same fixed parameter $\ka$. From the 5d perspective,  it is more 
natural to look for a prescription independent of any gauge 
fixing, and the most natural one is to simply add the 5d momenta 
$P_A$. This is similar to when we go from Newtonian mechanics to 
special relativity. Indeed we straightforwardly add the new 
relativistic 4-momenta. This does not simply reduce to a 
deformation of the 3-momentum addition since the addition of the 
(spatial) 3-momenta in special relativity involves explicitly the 
energies. In our context, let us define the total momentum in the 
center of mass frame as $R_A=P_A+Q_A$. Then the $\kappa$ 
parameter corresponding to the global two-particle system is
$$
\ka_{tot}^2=-R_AR^A=\ka_1^2+\ka_2^2-2P_AQ^A.
$$
In the special case when $P_A=Q_A$ and $\ka_1=\ka_2=\ka$, the 
total $\ka_{tot}$ is $2\ka$. This was actually already proposed 
in \cite{GL,GL2} as a possible solution to the soccer ball 
problem in DSR.

Another related issue regards the relative system for the two 
particles. More precisely, one can look for the Lorentz invariant 
distance between two DSR particles. This distance should involve 
the energy-momentum of both particles, the same way that the 
metric in general relativity would depend on the energy-momentum 
tensor of both particles. 

A more careful analysis would be required to understand the 
details of the separation of the two-particle system into global 
and relative systems.

\section{Discussion \label{sec_discussion}}

In this paper, we have studied a classical particle in five 
space-time dimensions subject to two constraints defining two 
energy scales $m$ and $\ka$. We have shown that, after gauge 
fixing, the 5d model can give rise to various DSR models in 4d. 
The reduction from 5d to 4d selects a set of phase space 
coordinates $(x,p)$ via the requirement that they should commute 
with both the $\ka$-shell constraint $\hh_{5d}$ and the gauge 
fixing function $\cc$. 

We found that Snyder's basis uses the 5d dilatation as its gauge 
fixing condition, while the bicrossproduct basis employs the 5d 
light-cone gauge. For both of these basis, we discussed the 
possible consequences of the scale $\ka$ on 4d physics. In 
principle, a ``natural'' gauge fixing $\cc=X_4-T$ could also be 
studied. From the 5d point of view, however, there is no clear 
reason to prefer one set of phase space coordinates over another 
one. This issue with coordinates is comparable to the one in 
General Relativity: General Relativity is diffeomorphism 
invariant and there is no canonical choice of ``physical 
coordinates,'' although we always use a particular coordinate 
system to do explicit computations and write down the physical 
predictions. 

In three space-time dimensions, the link between DSR and gravity 
has been clarified in \cite{Freidel:2005bb}. Indeed, in 3d 
quantum gravity, particles are identified as conical 
singularities and their momentum is defined through non-local 
measurements as (a function of) the holonomy around the particle. 
This explicit characterization allows to rigorously derive DSR 
from 3d quantum gravity and unambiguously compute the Feynman 
diagrams for the resulting non-commutative quantum field theory 
\cite{Freidel:2005bb}.

There is also a proposal attempting to move the similarity 
between DSR and GR to the level of an explicit relationship in 
four dimensions \cite{liberati}. In that proposal, the choice of 
coordinates $p_\mu$ (and $x_\mu)$ correspond to the definition of 
the measured momenta (and positions) in terms of the tetrad field 
$e_\mu^I$. The issue then becomes: what are we exactly measuring 
physically when we talk bout the energy-momentum $p_\mu$? The 
answer to this question will determine the ``correct'' choice of 
physical coordinates to use in DSR. Regardless, we expect the 
physical predictions of DSR to be independent of any gauge fixing 
choice and propose that the ``correctness'' of a particular 
choice of coordinates should be measured by how convenient these 
coordinates are to express the measurements of a particular 
observer. For instance, one could try to properly define length 
measurements using clocks and time-of-flight experiments to 
define the metric operationally.

At the end of the day, we cannot make concrete predictions using 
DSR as long as we do not find gauge invariant quantities 
(commuting with the two constraints of the 5d action) and their 
physical interpretation, or equivalently an explicit link between 
the choices of gauge fixing and measurement. This avenue of 
research seems to be a natural one from the 5d perspective. It is 
also our view that the 5d perspective should be a used when 
looking at two-particle systems and studying their properties. 
Other related topics to be investigated are free spinning 
particles.

Finally, an important unresolved issue regards the physical 
interpretation of the fifth dimension. Written as a 5d theory, 
DSR appears as a large extra dimension theory. We have proposed 
to see the coordinates in the fifth dimension as some effective 
degree of freedom coming from quantum gravity. The reformulation 
of GR as a $\SO(4,1)$ BF gauge field theory proposed in 
\cite{l&a} may prove to be a guide in this direction. It is also 
very tempting to interpret $P_4$ as the energy scale in a 
renormalisation scheme, as some kind of dynamical cut-off. $X_4$ 
would then be the generator of scale transformations. Such a 
speculation is supported by the fact that $X_4$ is (more or less) 
the 4d dilatation operator in the Snyder basis, but this is truly 
little evidence. One could look at the renormalisation equation 
of a scalar field and try to interpret them as equations of 
motion in the DSR framework. The potential link between DSR and 
quantum gravity and the fact that the renormalisation flow of 
general relativity can be associated to a fifth dimension (with 
an AdS metric) \cite{percacci} also points toward such an 
interpretation.

%%%%%%%%%%%%%%%%%%%%%%%
\section*{Acknowledgments}
FG and EL would like to thank Laurent Freidel for many 
explanations on the Dirac observables for the relativistic 
particle. JKG wants to thank Perimeter Institute for hospitality 
in September 2005 when part of this work was completed. For JKG 
this work is partially supported by the KBN grant 1 P03B 01828. 

%%%%%%%%%%%%%%%%%%%%%%%%%%%%%%%%%%%%%%%%%%%%%%%%%%%%%%%%%%%%%%%%%%%%%%%%%%
%%%%%%%%%%%%%%%%%%%%%%%%%%%%%%%%%%%%%%%%%%%%%%%%%%%%%%%%%%%%%%%%%%%%%%%%%%
%%%%%%%%%%%%%%%%%%%%%%%%%%%%%%%%%%%%%%%%%%%%%%%%%%%%%%%%%%%%%%%%%%%%%%%%%%
%%%%%%%%%%%%%%%%%%%%%%%%%%%%%%%%%%%%%%%%%%%%%%%%%%%%%%%%%%%%%%%%%%%%%%%%%%
%%%%%%%%%%%%%%%%%%%%%%%%%%%%%%%%%%%%%%%%%%%%%%%%%%%%%%%%%%%%%%%%%%%%%%%%%%

%%%%%%%%%%%%%%%%%%%%%%%%%%%%%%%%%%%%%%%%%%%%%%%%%%%%%%%%%%%%%%%%%%%%%%%%%%
%%%%%%%%%%%%%%%%%%%%%%%%%%%%%%%%%%%%%%%%%%%%%%%%%%%%%%%%%%%%%%%%%%%%%%%%%%
%%%%%%%%%%%%%%%%%%%%%%%%%%%%%%%%%%%%%%%%%%%%%%%%%%%%%%%%%%%%%%%%%%%%%%%%%%
%%%%%%%%%%%%%%%%%%%%%%%%%%%%%%%%%%%%%%%%%%%%%%%%%%%%%%%%%%%%%%%%%%%%%%%%%%
%%%%%%%%%%%%%%%%%%%%%%%%%%%%%%%%%%%%%%%%%%%%%%%%%%%%%%%%%%%%%%%%%%%%%%%%%%


\newpage

\begin{thebibliography}{99}

\bibitem{snyder}
H.~Snyder,
{\it Quantized Space-Time},
Phys.Rev. {\bf 71}, 38 (1947).

%\cite{Donkov:1984fj}
\bibitem{Donkov:1984fj}
  A.~D.~Donkov, R.~M.~Ibadov, V.~G.~Kadyshevsky, M.~D.~Mateev and M.~V.~Chizhov,
  {\it Quantum Field Theory And A New Universal High-Energy Scale: Dirac Fields},
  Nuovo Cim.\ A {\bf 87} (1985) 373.
  %%CITATION = NUCIA,A87,373;%%

%\cite{Donkov:1984ax}
\bibitem{Donkov:1984ax}
  A.~D.~Donkov, R.~M.~Ibadov, V.~G.~Kadyshevsky, M.~D.~Mateev and M.~V.~Chizhov,
  {\it Quantum Field Theory And A New Universal High-Energy Scale. Gauge Vector
  Fields},
  Nuovo Cim.\ A {\bf 87} (1985) 350.
  %%CITATION = NUCIA,A87,350;%%

%\cite{Kadyshevsky:1983yc}
\bibitem{Kadyshevsky:1983yc}
  V.~G.~Kadyshevsky and M.~D.~Mateev,
  {\it Quantum Field Theory And A New Universal High-Energy Scale: The Scalar
  Model},
  Nuovo Cim.\ A {\bf 87} (1985) 324.
  %%CITATION = NUCIA,A87,324;%%

\bibitem{majidbook} 
  S.~Majid. {\em Foundations of Quantum Groups Theory}, Cambridge University Press.

\bibitem{Matschull:1997du}
  H.~J.~Matschull and M.~Welling, 
  {\it Quantum mechanics of a point particle in 2+1 dimensional gravity},
  Class.\ Quant.\ Grav.\ {\bf 15} (1998) 2981 [arXiv:gr-qc/9708054].

\bibitem{Freidel:2005bb}
  L.~Freidel and E.~R.~Livine,
  {\it Ponzano-Regge model revisited. III: Feynman diagrams and effective field
  theory},
  arXiv:hep-th/0502106.

\bibitem{Imilkowska:2005vs}
  K.~Imilkowska and J.~Kowalski-Glikman,
  {\it Doubly special relativity as a limit of gravity},
  arXiv:gr-qc/0506084.

\bibitem{Amelino-Camelia:2000ge}
  G.~Amelino-Camelia, 
  {\it Testable scenario for relativity with minimum-length},
  Phys.\ Lett.\ B {\bf 510}, 255 (2001) [arXiv:hep-th/0012238].

\bibitem{Amelino-Camelia:2000mn}
  G.~Amelino-Camelia, 
  {\it Relativity in space-times with short-distance structure
  governed by an observer-independent (Planckian) length scale},
  Int.\ J.\ Mod.\ Phys.\ D {\bf 11}, 35 (2002) [arXiv:gr-qc/0012051].

\bibitem{jkgminl} 
  J.~Kowalski-Glikman,
{\it Observer independent quantum of mass},
  Phys.\ Lett.\ A {\bf 286} (2001) 391
  [arXiv:hep-th/0102098].

\bibitem{rbgacjkg} 
  N.~R.~Bruno, G.~Amelino-Camelia and J.~Kowalski-Glikman,
  {\it Deformed boost transformations that saturate at the Planck scale},
  Phys.\ Lett.\ B {\bf 522} (2001) 133 [arXiv:hep-th/0107039].

\bibitem{Magueijo:2001cr}
  J.~Magueijo and L.~Smolin, 
  {\it Lorentz invariance with an invariant energy scale},
  Phys.\ Rev.\ Lett.\  {\bf 88} (2002) 190403 [arXiv:hep-th/0112090].

\bibitem{Kowalski-Glikman:2002ft}
  J.~Kowalski-Glikman,
  {\it De Sitter space as an arena for doubly special relativity},
  Phys.\ Lett.\ B {\bf 547} (2002) 291
  [arXiv:hep-th/0207279].
  
\bibitem{Kowalski-Glikman:2003we}
  J.~Kowalski-Glikman and S.~Nowak,
  {\it Doubly special relativity and de Sitter space},
  Class.\ Quant.\ Grav.\  {\bf 20} (2003) 4799
  [arXiv:hep-th/0304101].

\bibitem{Blaut:2003wg}
  A.~Blaut, M.~Daszkiewicz, J.~Kowalski-Glikman and S.~Nowak,
  {\it Phase spaces of doubly special relativity},
  Phys.\ Lett.\ B {\bf 582} (2004) 82
  [arXiv:hep-th/0312045].
 

\bibitem{Kowalski-Glikman:2001px}
  J.~Kowalski-Glikman,
  {\it Planck-scale relativity from quantum kappa-Poincare algebra},
  Mod.\ Phys.\ Lett.\ A {\bf 17}, 1 (2002)
  [arXiv:hep-th/0107054].

\bibitem{FGL}
L.~Freidel, F.~Girelli and E.~R.~Livine,
{\it Dirac observables and the Algebra of Deformed Special Relativity},
in preparation

\bibitem{GL}
F.~Girelli and E.~R.~Livine,
{\it Physics of Deformed Special Relativity: Relativity Principle revisited},
[arXiv:gr-qc/0412004].

\bibitem{GL2}
F.~Girelli and E.~R.~Livine,
{\it Physics of Deformed Special Relativity},
Braz.J.Phys. {\bf 35} (2005) 432-438,
[arXiv:gr-qc/0412079].

\bibitem{l&a}
L.~Freidel and A.~Starodubtsev,
{\it Quantum gravity in terms of topological observables},
[arXiv:hep-th/0501191].

\bibitem{rainbow}
J.~Magueijo and L.~Smolin,
{\it Gravity's Rainbow},
Class.Quant.Grav. 21 (2004) 1725-1736,
[arXiv:gr-qc/0305055].

\bibitem{DL}
E.~R.~Livine and D.~Oriti,
{\it About Lorentz invariance in a discrete quantum setting},
JHEP 0406 (2004) 050,
[arXiv:gr-qc/0405085].

\bibitem{ghosh}
S.~Ghosh, {\it The $AdS$ particle},  Phys.Lett. B623 (2005) 251-257, [arXiv:hep-th/0506084].


\bibitem{ruegg}
S.~Majid and H.~Ruegg, {\it Bicrossproduct structure of kappa Poincare group and noncommutative geometry}, Phys.\ Lett.\ B {\bf 334} (1994) 348, [arXiv:hep-th/9405107]; \\
J.~Lukierski, H.~Ruegg and W.~J.~Zakrzewski, {\it Classical quantum mechanics of free kappa relativistic systems}, Annals Phys.\  {\bf 243} (1995) 90, [arXiv:hep-th/9312153].

%\bibitem{VilelaMendes:2004up}
\bibitem{mendes}
R.~Vilela Mendes,
{\it Some consequences of a noncommutative space-time structure},
Eur.\ Phys.\ J.\ C {\bf 42}, 445 (2005)
[arXiv:hep-th/0406013].

\bibitem{liberati}
R.~Aloisio, A.~Galante, A.~Grillo, S.~Liberati, E.~Luzio and F.~Mendez,
{\it Deformed Special Relativity as an effective theory of measurements on quantum gravitational backgrounds},
[arXiv:gr-qc/0511031].

\bibitem{york}
J.~D.~Brown and J.~W.~York,
{\it Quasilocal energy and conserved charges derived from the gravitational action},
Phys.Rev.D {\bf 47}, 4 (1993).

\bibitem{percacci}
R.~Percacci,
{\it The Renormalization Group, Systems of Units and the Hierarchy Problem},
[arXiv: hep-th/0409199].

\end{thebibliography}
\end{document}